\documentclass[aps,prl,reprint,twocolumn,amsmath,amssymb,showpacs,superscriptaddress,floatfix]{revtex4-2}

\usepackage{color}
\usepackage{textcomp} 
\usepackage{mathtools}
\usepackage{mathrsfs,amsmath}
\usepackage{graphicx}
\graphicspath{{Figures/}} 
\usepackage{lipsum}
\usepackage{dcolumn}
\usepackage[mathlines]{lineno}
\usepackage{hyperref}
\usepackage{xcolor}
\hypersetup{
    colorlinks,
    linkcolor={red!50!black},
    citecolor={blue!50!black},
    urlcolor={blue!80!black}
}
\usepackage{bm}
\usepackage{epstopdf}
\usepackage{soul}
\usepackage{epsfig}
\usepackage{bbold}
\usepackage{braket}
\usepackage{physics}
\usepackage{gensymb}
\usepackage{amsmath,amssymb}
\usepackage{bigints} 

\begin{document}

\title{Generalized Polarization Matrix Approach to Near-Field Optical Chirality}
\author{Kayn~A.~Forbes}
\email{K.Forbes@uea.ac.uk}
\author{David L. Andrews}

\affiliation{School of Chemistry, University of East Anglia, Norwich Research Park, Norwich NR4 7TJ, United Kingdom}

\begin{abstract}

For paraxial light beams and electromagnetic fields, the Stokes vector and polarization matrix provide equivalent scalar measures of optical chirality, widely used in linear optics. However, growing interest in non-paraxial fields, with fully three-dimensional polarization components, necessitates an extended framework. Here, we develop a general theory for characterizing optical chirality in arbitrary electromagnetic fields, formulated through extensions of the polarization matrix approach. This framework applies to both near- and far-field optical helicity and chirality. As examples, we demonstrate its relevance to near-zone fields from chiral dipole emission and the focal plane of tightly focused beams. 

\end{abstract}

\maketitle

\section{Introduction}
The polarization of light has long been a key aspect of research in optics.  Since the pioneering work of Christiaan Huygens, and through its long subsequent development, copious theoretical methods have been devised to describe and quantify optical polarization \cite{born2013principles, mandel1995optical, wolf2007introduction, brosseau1998fundamentals, goodman2015statistical}. Today, understanding and manipulating the associated electromagnetic field vectors plays a critical role in modern optics and photonics \cite{saleh2019fundamentals, chekhova2021polarization}. Widespread theoretical representations include the Jones vector and associated matrices, the Stokes vector, and Mueller matrices. Amongst more recent developments, Hermitian polarization density matrices represent not only the polarization properties of light, but also coherence. Emil Wolf was one of the principal proponents of such matrices to describe the coherence and polarization properties of light, and their interplay \cite{mandel1995optical, wolf2007introduction}. 

Freely propagating electromagnetic fields in vacuum may be characterized by a multiplicity of conserved quantities \cite{fushchich1992complete, bliokh2013dual}. Amongst them, one conserved quantity of special interest is the optical helicity \cite{barnett2012duplex, cameron2012optical, aiello2022helicity}. It emerges that this parameter is directly engaged in any optical measurement that can elicit the subtle differences between the properties of chiral materials with opposite handedness. The optical helicity is a pseudoscalar, which can be defined in terms of the projection of spin angular momentum onto the canonical momentum density. Closely related, and directly proportional to it for monochromatic fields \cite{mackinnon2019differences}, is the optical chirality \cite{bliokh2011characterizing}. For paraxial fields, optical helicity is itself directly proportional to the degree of polarization ellipticity, taking its maximum value for circular polarizations, while the helicity for linearly polarized or unpolarized paraxial light is zero.  

In any paraxial formulation, the consistency of the transverse beam profile allows optical helicity to be quantified in a 2D polarization basis.  In this sense, plane waves are a classic example of so-called ‘2D light’: they are polarized only in the plane transverse to the beam propagation.  In modern nano-optics, however, where light is spatially confined to wavelength scales, such a description is insufficient to account for the distinctly different behavior that is observed, due to longitudinal field components along the direction of propagation. Such non-paraxial, or 3D,  structured light therefore requires a different approach for a full characterization of its optical properties, necessitating extensions of the 2D polarization matrix or Stokes vector approach to 3D \cite{carozzi2000parameters, setala2002degree, dennis2004geometric, sheppard2014jones, brasselet2023polarization, alonso2023geometric, herrera2024stokes}. 

It was previously noted that the third Stokes parameter $S_3$, based on a 2D polarization approach, is not a suitable measure of near-field optical chirality \cite{leeder2015point}. Here we secure general theory applicable to arbitrary electromagnetic fields, allowing the optical helicity (chirality) to be quantified through either a polarization matrix or Stokes vector approach. Our theory highlights how non-paraxality of electromagnetic fields necessitates a description based on mixed electric-magnetic 3D polarization matrices. We highlight applications of our general theory via the specific cases of chiral emission, and tightly-focused chiral beam optics. 

\section{OPTICAL HELICITY AND OPTICAL CHIRALITY}





In its most general form, the optical helicity density – a conserved quantity for arbitrary electromagnetic fields in vacuum – is given by \cite{cameron2012optical, aiello2022helicity, bliokh2013dual}:
\begin{align}
    h = \frac{\epsilon_0 c}{2} \left( \mathbf{A}^{\perp} \cdot \mathbf{B} - \mathbf{C}^{\perp} \cdot \mathbf{E} \right),
    \label{eq:1}
\end{align}
where $\mathbf{A}^{\perp}$ and $\mathbf{C}^{\perp}$ are transverse electromagnetic vector potentials, and $\mathbf{E}$ and $\mathbf{B}$ are the corresponding electric and magnetic fields. These quantities are related via the dual-symmetric definitions: $\mathbf{E} = -\nabla \times \mathbf{C}^{\perp} = -\dot{\mathbf{A}}^{\perp}$ and $\mathbf{B} = \nabla \times \mathbf{A}^{\perp} = -\dot{\mathbf{C}}^{\perp}$. For quasi-monochromatic fields – our focus here – the vector potentials are assumed to have harmonic time dependence, i.e., $\mathbf{A}^{\perp}(\mathbf{r}, t) = \mathrm{Re} \left\{ \mathbf{A}^{\perp}(\mathbf{r}) \text{e}^{-i \omega t} \right\}$, $\mathbf{C}^{\perp}(\mathbf{r}, t) = \mathrm{Re} \left\{ \mathbf{C}^{\perp}(\mathbf{r}) \text{e}^{-i \omega t} \right\}$, with complex phasors $\mathbf{A}^{\perp}(\mathbf{r})$ and $\mathbf{C}^{\perp}(\mathbf{r})$. This implies $i \omega \mathbf{A}^{\perp} = \mathbf{E}$, $i \omega \mathbf{C}^{\perp} = \mathbf{B}$, so Eq.~\eqref{eq:1} becomes:
\begin{align}
    h = -\frac{\epsilon_0 c}{2 \omega} \, \mathrm{Im} \left( \mathbf{E}^* \cdot \mathbf{B} \right),
    \label{eq:2}
\end{align}
where \( \omega \) is the central angular frequency. In this regime, the optical helicity density \( h \) is directly proportional to the optical chirality $\chi$:
\begin{align}
    \chi=-\frac{\epsilon_0\omega}{2}\text{Im}\bigl(\mathbf{E}^*\cdot\mathbf{B}\bigr)=\frac{\omega^2}{c}h. 
    \label{eq:3}
\end{align}
For a circularly polarized plane wave we have $\mathbf{E}=E_02^{-1/2}\bigl(\mathbf{\hat{x}}+i\sigma\mathbf{\hat{y}}\bigr)\text{exp}(i\mathbf{k}\cdot\mathbf{r})$ and $\mathbf{B}=c^{-1}E_02^{-1/2}\bigl(\mathbf{\hat{y}}-i\sigma\mathbf{\hat{x}}\bigr)\text{exp}(i\mathbf{k}\cdot\mathbf{r})$.   Inserting into Eq.~\eqref{eq:2} gives the following, well-known expression;
 
\begin{align}
    &h=\frac{I\sigma}{c\omega},
    \label{eq:4}
\end{align} 		
where $I$ is the intensity of the beam and $\sigma=\pm1$ for left(right) handed circular polarization; $\chi$ is easily determined from Eq.~\eqref{eq:4} using Eq.~\eqref{eq:3}.

For any plane wave, the orthogonality of the electric and magnetic fields, whose vector product $\mathbf{E}\times\mathbf{B}$ generates the Poynting vector in the axial direction, allows a facility to describe most electromagnetic fields purely in terms of the electric field, and thus by the electric polarization matrices and Stokes parameters.  These include the energy density, spin angular momentum density (including transverse spin), canonical momentum density, and orbital momentum density. While an involvement of pure magnetic polarization matrices may be necessary under certain circumstances, we show here that the optical helicity density for non-paraxial fields necessitates a formulation that entails the \textit{mixed} electric-magnetic tensors.  

\section{TWO-DIMENSIONAL DESCRIPTION} \label{III}

Let us consider a quasi-monochromatic paraxial (or plane wave) propagating from a source along the $z$ axis, with the transverse plane ($x, y$) also represented in Cartesian form.  The (electric) polarization matrix may be written as \cite{wolf2007introduction}: 

\begin{align}
\Phi_{ij}^{\text{E(2D)}}&=\epsilon_0c\langle E_i^*E_j\rangle,\quad i,j=x,y \nonumber \\ &=\epsilon_0c\begin{bmatrix}
\langle E_x^*E_x\rangle & \langle E_x^*E_y\rangle \\
\langle E_y^*E_x\rangle & \langle E_y^*E_y\rangle
\end{bmatrix},
\label{eq:5}
\end{align}
where angular brackets represent temporal averaging and the symbol $*$ denotes the complex conjugate. The Stokes four-vector, with components $S_i$, is a more direct representation of polarization information that can be inferred or extracted from intensity measurements. The polarization matrix Eq.~\eqref{eq:5} may be written in terms of the Stokes parameters $S_i$ as \cite{wolf2007introduction}:

\begin{align}
\Phi_{ij}^{\text{E(2D)}}&=\frac{1}{2}\begin{bmatrix}
S_0+S_1 & S_2+iS_3 \\
S_2-iS_3 & S_0-S_1
\end{bmatrix}.
\label{eq:6}
\end{align} 
The third Stokes parameter $S_3$ (written as $s_3$ when normalized to the total intensity, i.e. $s_3=S_3/S_0$) is often associated with measures of optical helicity and spin angular momentum density \cite{berestetskii2012quantum, bliokh2015transverse, leeder2015point, eismann2021transverse, alonso2023geometric, forbes2024orbit}. It represents the difference in intensities measured for right-circularly polarized and left-circularly polarized light:

\begin{align}
S_3=i\bigl(\Phi_{yx}^{\text{E}}-\Phi_{xy}^{\text{E}}\bigr)=\epsilon_0c \text{Im}\bigl[ \langle E_x^*E_y\rangle-\langle E_y^*E_x\rangle\bigr].
\label{eq:7}
\end{align}
On the condition that the trace of Eq.~\eqref{eq:5} gives the intensity of the beam $\text{Tr}\Phi_{ij}^{\text{E(2D)}}=I$, for a circularly polarized monochromatic plane wave the polarization matrix is readily determined as \cite{born2013principles}
 
\begin{align}
\Phi^\text{E(2D)}=\frac{I}{2}\begin{bmatrix}
1 & i\sigma \\
-i\sigma & 1
\end{bmatrix}.
\label{eq:8}
\end{align}
Inserting Eq.~\eqref{eq:8} into Eq.~\eqref{eq:7} gives
\begin{align}
S_3=I\sigma\label{eq:9}
\end{align}
Both $S_3$ and $h$ are pseudoscalar measures. While $S_3\propto h$, the physical dimensions of the Stokes vector are $\text{[M]} \text{[T]}^{-3}$, those of the beam helicity density are $\text{[M]} \text{[L]}^{-1} \text{[T]}^{-1}$. Optical chirality density $\chi$ is equally a pseudoscalar but has units of $\text{[M] [L]}^{-2}\text{[T]}^{-2}$.  Optical helicity couples to the mixed electric-magnetic dipole responses (E1M1) of chiral materials \cite{cameron2017chirality}: in contrast, all the Stokes parameters are intensities which engage with purely electric dipole responses E1E1 of a material detector. Thus, $S_3$ and optical helicity (chirality) are not identical and engage through different transitions in light-matter interactions. Nonetheless, in the paraxial (or far-field regime) there is a direct one-to-one mapping of the measurement of $S_3$ and the optical helicity, spin, and circular polarization of the beam measured. 

Using a purely electric polarization matrix Eq.~\eqref{eq:5} to describe the optical helicity of paraxial light, a quantity that explicitly involves the product of electric and magnetic fields (Eq.~\eqref{eq:2}), is only possible because in 2D polarized light (the paraxial approximation, i.e. $E_z=0$, $B_z=0$), then $E(\mathbf{r})=cB(\mathbf{r})$ and the scalar spatial distributions of the electric and magnetic fields are equivalent. This is because $c\mathbf{B}=\mathbf{\hat{z}}\times\mathbf{E}$ for paraxial and plane wave light. However, this is not true in general, and for non-paraxial light the electric and magnetic fields are not locally locked with each other \cite{berry2009optical, bliokh2013dual}, we discuss the implications of this in detail in the next section.

\section{THREE-DIMENSIONAL DESCRIPTION}

In this Section we will show that there is no corresponding 3D Stokes parameter or 3D polarization matrix, based purely on electric fields, that can describe the optical helicity of a 3D structured light field, i.e., an arbitrary electromagnetic field.  It is important to reiterate that 3D polarization is principally a near-field phenomenon, whilst standard 2D Stokes theory is widely applied to infer 3D polarization structure in far-field measurements – see for example \cite{nechayev2019orbital}. 

First, it is instructive to show that the pure electric polarization matrix does not deliver what we require.  For 3D-polarized fields the electric polarization matrix elements can be written as follows, now including the non-zero longitudinal components for non-paraxial fields:
\begin{align}
\Phi_{ij}^{\text{E(3D)}}&=\epsilon_0c\langle E_i^*E_j\rangle,\quad i,j=x,y,z \nonumber \\ &=\epsilon_0c\begin{bmatrix}
\langle E_x^*E_x\rangle & \langle E_x^*E_y\rangle & \langle E_x^*E_z\rangle \\
\langle E_y^*E_x\rangle & \langle E_y^*E_y\rangle & \langle E_y^*E_z\rangle \\
\langle E_z^*E_x\rangle & \langle E_z^*E_y\rangle & \langle E_z^*E_z\rangle
\end{bmatrix}.
\label{eq:10}
\end{align}
Noting that for non-paraxial fields $E(\mathbf{r})\neq cB(\mathbf{r})$ is it obvious that the polarization matrix Eq.~\eqref{eq:10} cannot describe the optical helicity (or chirality) of non-paraxial electromagnetic fields since $E_z(\mathbf{r})\neq cB_z(\mathbf{r})$. That is, in contrast to the transverse electric and magnetic fields, the spatial distributions of the longitudinal electric and magnetic fields are not the same. As the optical chirality (helicity) Eq.~\eqref{eq:3} entails the inner product of the electric and magnetic fields, it will in general have non-zero contributions from $E_z^*B_z$.

In order to describe optical chirality (helicity) it emerges that, for a more general case, we can utilize the following equation:
\begin{align}
h=\frac{\text{Im}\delta_{ij}}{2\omega}\bigl(\Phi_{ij}^\text{N}-\Phi_{ij}^\text{M}\bigr),
\label{eq:11}
\end{align}
where the summation is implied of the terms with repeated indices. In Eq.~\eqref{eq:11}, in order to account for non-paraxial optical helicity of 3D light, we introduce mixed electric-magnetic polarization tensors:
\begin{align}
\begin{rcases}
&\Phi_{ij}^{\text{M(3D)}}=\epsilon_0c\langle E_i^* B_j \rangle, \\ &\Phi_{ij}^{\text{N(3D)}}=\epsilon_0c\langle B_i^* E_j \rangle 
\quad\end{rcases}  i,j=x,y,z. 
\label{eq:12}
\end{align} 
Eq.~\eqref{eq:11} represents the key result of this work. It is a measure of optical helicity (chirality) that is applicable to both near- and far-field electromagnetic fields.  Duly taking account all 3D components of both electric and magnetic field components enables the optical helicity density to be calculated for arbitrary electromagnetic fields. As explicit proof of this approach to measure the optical helicity and chirality densities of electromagnetic fields we now apply these results to two important near-field chiral phenomena. 

\section{CHIRAL DIPOLE RADIATION}

In work by Leeder et al. \cite{leeder2015point}, duly appropriating retarded fields to study the range-dependence of emission from a chiral dipole, it was correctly concluded that $S_3$ is not in any sense a viable measure of optical chirality in the near-field. With the present, general formulation for optical helicity of arbitrary fields, we now show how the method described above also works when cast in terms of the corresponding field propagation tensors.  This enables results to be secured for the optical helicity of radiated fields over arbitrary distances from a chiral source. Radiative emission from any such source will engage electronic transitions that simultaneously satisfy both electric and magnetic dipole selection rules. In consequence the radiation has electric and magnetic components each partially dependent on both kinds of transition moment. Using the implied summation convention for repeated tensor indices, the electromagnetic field vectors are expressible in terms of the oscillating electric and magnetic transition dipoles, $\boldsymbol{\mu}$ and $\mathbf{m}$ respectively, as follows: 
\begin{align}
E_j=-\mu_iV_{ij}-c^{-1}m_iU_{ij}
\label{eq:13}
\end{align}
\begin{align}
B_j=-\mu_iU_{ij}-c^{-1}m_iV_{ij}
\label{eq:14}
\end{align}
where the Green’s function field propagation tensors $V_{ij}$ and $U_{ij}$ are defined in terms of wavenumber $k$, and the displacement vector $\mathbf{R}$ from the source \cite{zangwill2013modern, salam2009molecular}:
\begin{align}
V_{ij}&=\frac{\text{e}^{ikR}}{4\pi\epsilon_0R^3}
\Bigl[\Bigl(1-ikR\Bigr)\Bigl(\delta_{ij}-3\hat{R}_i\hat{R}_j\Bigr)\nonumber \\ & - k^2R^2\Bigl(\delta_{ij}-\hat{R}_i\hat{R}_j\Bigr)\Bigr]\label{eq:15}
\end{align}
\begin{align}
U_{ij}=\frac{\text{e}^{ikR}}{4\pi\epsilon_0R^3}\epsilon_{ijk}{\hat{R}_k}
\Bigl[ikR+k^2R^2\Bigr]
\label{eq:16}
\end{align}
Interestingly the element of the electric field given by the first term on the right of Eq.~\eqref{eq:13}, associated with an electric dipole character of the emission, contains terms of both transverse and longitudinal aspect with respect to the vectorial displacement from the source – the longitudinal part associated with the initial term in Eq.~\eqref{eq:15} for the propagator $V_{ij}$. However, the element of the electric field associated with magnetic dipole character is purely transverse with respect to $\mathbf{R}$ since the second term in Eq.~\eqref{eq:13} arises from the propagator $U_{ij}$ Eq.~\eqref{eq:16} which includes the Levi-Civita antisymmetric tensor $\epsilon_{ijk}$ (thereby entailing the unit vector $\mathbf{\hat{R}}$ in a vector cross-product). Previous studies have usually been only concerned with the electric field component Eq.~\eqref{eq:13} - but in the near-zone one cannot neglect Eq.~\eqref{eq:14} as its longitudinal components will differ from those of Eq.~\eqref{eq:13}, as we have discussed. For both propagation tensors, there is a striking difference in their forms of amplitude and phase evolution, for the transverse and longitudinal components: the detailed time and space evolution are described in detail in \cite{rice2012identifying}. 

The electric-magnetic polarization matrices Eq.~\eqref{eq:12} that describe Eq.~\eqref{eq:13} and Eq.~\eqref{eq:14} are too large to reproduce here, but after some significant calculations we arrive at:
\begin{align}
h_{\text{dipole}} &= \frac{\text{Im}\delta_{ij}}{2\omega}\bigl(\Phi_{ij}^\text{N}-\Phi_{ij}^\text{M}\bigr) \nonumber \\ &= -\frac{\mu_0c}{(4\pi)^2\omega}\text{Im}\Bigl[\frac{k^4}{R^2}\Bigl(\mu_x m_x + \mu_y m_y\Bigr) \nonumber \\ & + 4\mu_z m_z\Bigl(\frac{1}{R^6}+\frac{k^2}{R^4}\Bigr)
\Bigr].
\label{eq:17}
\end{align}
This is to the best of our knowledge the first time the optical helicity (chirality) density of an oscillating chiral dipole has been produced that is applicable to all distances from the source. For distances of around and above an optical wavelength, the full expression of equation Eq.~\eqref{eq:17} shows the emerging retardation character of optical propagation. Analogous with the unified theory of resonance energy transfer (RET) \cite{andrews1989unified, daniels2003resonance}, the `unified theory' of chiral dipolar emission Eq.~\ref{eq:17} displays not only a near-zone $R^{-6}$ and far-zone $R^{-2}$ distance-dependence, it also has an $R^{-4}$ intermediate distance dependence. 

In the long-range far-zone (FZ), where $R$ is much larger than the wavelength, we secure the following asymptotic result, consistent with paraxial 2D polarization:
\begin{align}
h(\text{FZ}) &= -\frac{\mu_0c}{(4\pi)^2\omega}\frac{k^4}{R^2}\text{Im}\Bigl[\mu_x m_x + \mu_y m_y\Bigr],
\label{eq:18}
\end{align}
the anticipated inverse-square dependence on distance matching the result of Leeder et al.  [12].  

Conversely, for the near-zone (NZ) limit applicable close to the source, where $kR<<1$ is much less than unity, we retrieve:
\begin{align}
h(\text{NZ}) &= -\frac{\mu_0c}{(4\pi)^2\omega}\frac{1}{R^6}\text{Im} 4\mu_z m_z.
\label{eq:19}
\end{align}
The result Eq.~\eqref{eq:19} thus represents the conveyance of optical helicity over sub-wavelength dimensions. The helicity density at short distances from a chiral source is a subject that itself may have significant analytical applications for nanoscale sensing as, for example, in studying the transmission of a structured beam focused upon a planar medium \cite{nechayev2021kelvin} and the near-field optical chirality of plasmonic nanostructures \cite{kakkar2020superchiral, hu2022plasmonic}.

The aforementioned work by Leeder et al. \cite{leeder2015point} focused on the electric field contributions to the emission Eq.~\eqref{eq:13}, consistent with the conventional form of the Stokes parameter representation. As was shown, whereas the Stokes vector can adequately  portray optical helicity far from a source of chiral emission, it gives an incorrect picture in the near-zone. However, the inclusion of the magnetic field contributions from Eq.~\eqref{eq:14} and using Eq.~\eqref{eq:11} now enables results to be secured that fully describe the near-field (and intermediate field) chirality, including terms that originate solely from the longitudinal components of the transition moments, $\mu_{z}$ and $m_{z}$ as highlighted in Eq.~\eqref{eq:17}. 

An additional insight is afforded by considering this result as the time inverse of light converging to a focus. We now consider its application as a direct measure of helicity conveyance in the non-paraxial regime within the focal region of a focused laser beam.  

\section{Focused Vortex Optics}

Optical vortices are beams of light that have helical wavefronts described by the multiplier $\text{e}^{i\ell\phi}$,
where $\phi$ is the polar angle in the beam transverse plane, and $\ell\in\mathbb{Z}$ is the pseudoscalar topological charge. These helical wavefronts engender such beams with optical orbital angular momentum (OAM) of $\ell\hbar$ along the direction of propagation. There has been significant interest in whether geometrically chiral optical vortex beams with helical wavefronts, $\ell>0$ being left-handed, $\ell<0$ right-handed, could engage in chiral light-matter interactions and optical activity. For a historical perspective and review of this exciting field see \cite{forbes2021orbital, porfirev2023light}. It is now well understood that 2D linearly polarized tightly focused optical vortex beams possess optical helicity density that is independent of the 2D polarization ellipticity \cite{forbes2022optical, green2023optical, forbes2023customized}.  Consider for example an $x$-polarized Laguerre Gaussian beam which includes a longitudinal component.  The electric polarization matrix emerges as follows, in which ($r$, 
$\phi$) are the transverse polar coordinates (radial distance and azimuthal angle) and $\gamma$ is defined as $\gamma = \frac{|\ell|}{r}-\frac{2r}{w^2}+ \frac{ikr}{R[z]} -\frac{4r}{w^2}\frac{L^{|\ell|+1}_{p-1}}{L^{|\ell|}_{p}}$ (see \cite{green2023optical} for details):
\begin{widetext}
\begin{align}
\Phi_{ij}^{\text{E(3D)}}&=I\begin{bmatrix}
1 & 0 & \frac{i}{k}\Bigl(\gamma\cos\phi-\frac{i\ell}{r}\sin\phi\Bigr) \\
0 & 0 & 0 \\
-\frac{i}{k}\Bigl(\gamma\cos\phi+\frac{i\ell}{r}\sin\phi\Bigr) & 0 & \frac{1}{k^2}\Bigl(|\gamma|^2\cos^2\phi+\frac{\ell^2}{r^2}\sin^2\phi\Bigr)
\end{bmatrix}.
\label{eq:20}
\end{align}
\end{widetext}
The optical helicity density for this field, experimentally verified \cite{wozniak2019interaction, rouxel2022hard}, takes the form \cite{forbes2021measures, green2023optical}:
\begin{align}
h=-\frac{I_\text{LG}}{c\omega}\frac{\ell}{k^2r}\gamma.
\label{eq:21}
\end{align}
Note this helicity (chirality) density is completely independent of $\sigma$, i.e. the state of circular polarization. From the pure electric polarization matrix Eq.~\eqref{eq:20} there is insufficient information to extract the optical helicity density Eq.\eqref{eq:21}.  However, the corresponding mixed electric-magnetic matrix $\Phi_{ij}^{\text{M(3D)}}$ (Eq.\eqref{eq:12}) is cast as follows:
\begin{widetext}
\begin{align}
\Phi_{ij}^{\text{M(3D)}}&=\frac{I}{c}\begin{bmatrix}
0 & 0 & \frac{i}{k}\Bigl(\gamma\sin\phi+\frac{i\ell}{r}\cos\phi\Bigr) \\
0 & 0 & 0 \\
0 & -\frac{i}{k}\Bigl(\gamma^*\cos\phi+\frac{i\ell}{r}\sin\phi\Bigr) & \frac{1}{k^2}\Bigl(\frac{i\ell}{r}\gamma+\Bigl\{|\gamma|^2-\frac{\ell^2}{r^2}\Bigr\}\cos\phi\sin\phi\Bigr)
\end{bmatrix}
\label{eq:22}
\end{align}
\end{widetext}
Then, using $\Phi_{ij}^{\text{M}}=\Phi_{ji}^{*\text{N}}$ and inserting into Eq.~\eqref{eq:11} gives the correct helicity of Eq.~\eqref{eq:21}. 

\section{THREE-DIMENSIONAL STOKES PARAMETER APPROACH}

As we saw in Section~\ref{III}, the 2D polarization matrix Eq.~\eqref{eq:5} approach is equivalent to an alternative description framed in terms of the four real Stokes vectors $S_0$, $S_1$, $S_2$, and $S_3$. Similarly, 3D polarization matrix formulations have a counterpart in terms of nine real (3D) Stokes parameters $\Lambda_i$ \cite{sheppard2014jones}. Here we briefly outline the approach of measures of optical chirality through the 3D Stokes vector, applicable to both near- and far-fields, as an alternative to the 3D polarization matrix we have introduced in this work.  
The 9 individual expressions for $\Lambda_i$, and details of how they relate to the components of the 3D polarization matrixes, can be found in \cite{sheppard2014jones}, we only explicitly require $\Lambda_0=\Phi_{xx}+\Phi_{yy}+\Phi_{zz}$. The electric 3D Stokes parameter which plays an analogous role to $S_3$ is $\Lambda_{2}^{\text{E}}=\frac{3}{2}i\bigl(\Phi_{xy}^{\text{E(3D)}}-\Phi_{yx}^{\text{E(3D)}}\bigr)$: this 3D Stokes parameter, using Eq.~\eqref{eq:20},  will clearly not generate the optical helicity Eq.~\eqref{eq:21}, for example, for reasons we have established. The correct way to apply the 3D stokes parameter is to formulate it as follows:
\begin{align}
h=\frac{\text{Im}\bigl(\Lambda_{0}^{\text{N}}-\Lambda_{0}^{\text{M}}\bigr)}{2\omega},
\label{eq:23}
\end{align}
where $\Lambda_{0}^{\text{N(M)}}=\Phi_{xx}^{\text{N(M)}}+\Phi_{yy}^{\text{N(M)}}+\Phi_{zz}^{\text{N(M)}}$ are the elements of the mixed electric-magnetic matrices Eq.~\eqref{eq:12}. The measure of chirality based on 3D Stokes parameters Eq.~\eqref{eq:23} gives equivalent results to expression Eq.~\eqref{eq:11} which is based on a polarization matrix approach.

\section{Discussion and CONCLUSION}

It was remarked in \cite{leeder2015point} that the third Stokes parameter $S_3$ was not a suitable measure of optical chirality in the near-field. We have now provided the general theory to calculate the optical helicity density using both polarization matrices and Stokes vector representations for arbitrary electromagnetic fields. Crucial to our method is recognizing that in non-paraxial fields $E(\mathbf{r})\neq B(\mathbf{r})$ and we thus require the use of mixed electric and magnetic polarization matrices Eq.~\eqref{eq:12}. A key insight that this theory provides is that, for arbitrary electromagnetic fields, optical helicity (chirality) cannot be directly identified with either spin or circular polarization (or a degree of ellipticity) \cite{forbes2024orbit}. The longitudinally polarized fields produce optical chirality and helicity, and of course these cannot be circularly polarized as they both oscillate along the same single axis. In the far field, optical helicity (chirality) is unequivocally proportional to the circularity of the 2D polarization state.  However, in the near-field (and therefore in general) circularity in the polarization of the light is not required for optical helicity (chirality). 

For 3D structured light, the electric polarization matrix cannot describe conserved electromagnetic field properties in general, such as the energy density and spin angular momentum density. Both of these for example have magnetic contributions due to the dual symmetry of Maxwell's equation \cite{bliokh2013dual}, and thus their formulation requires the inclusion of a pure magnetic polarization tensor $\Phi^{\text{B(3D)}}_{ij}=\langle B^*_iB_j\rangle \quad i,j=x,y,z$. Moreover, the whole dual-symmetric nature of the optical properties of 3D polarized light could be completely described by a single $6\times6$ electric-magnetic matrix if desired. What is crucially important with optical helicity and chirality is that it engages \text{both} the electric and magnetic transitions of the material. Although non-paraxial fields have different distributions of energy and spin for the electric and magnetic parts, for example, due to $E_z(\mathbf{r})\neq cB_z(\mathbf{r})$, in experiments we generally measure the former, due to the electric-biased (dual asymmetric) nature of most materials. As such, the electric polarization matrix is usually sufficient for the characterization of these quantities. We have shown here that this does not extend to optical helicity or chirality and that we require the use of \textit{mixed} electric-magnetic descriptions Eqs.~\eqref{eq:11}, \eqref{eq:12}, and \eqref{eq:23}. We applied this general theory successfully to two specific examples: chiral dipolar emission and tightly focused optical vortex beams. The former resulted in a unified theory of chiral emission from a dipolar source, which to the best of our knowledge is the first time this has been shown in the literature.

\bibliography{references.bib}

\end{document}